\documentclass{aa}

\usepackage{psfig}
\usepackage{graphicx}
\usepackage{natbib}
\usepackage[english]{babel}  
\usepackage{txfonts}
\usepackage{url}

\begin{document}

\title{Thermal emission, shock modification, and X-ray emitting ejecta in SN~1006}

\author{M. Miceli\inst{1,2,3} \and F. Bocchino\inst{2,3} \and D.~Iakubovskyi\inst{4} \and S. Orlando\inst{2,3} \and I. Telezhinsky\inst{5} \and M. G. F. Kirsch\inst{6} \and O. Petruk\inst{3,7,8} \and  G. Dubner\inst{9} \and G. Castelletti\inst{9} }

\offprints{M. Miceli,\\ \email{miceli@astropa.unipa.it}}

\institute{Dipartimento di Scienze Fisiche ed Astronomiche, Sezione di 
Astronomia, Universit{\`a} di Palermo, Piazza del Parlamento 1, 90134 Palermo, 
Italy
\and
INAF - Osservatorio Astronomico di Palermo, Piazza del Parlamento 1, 90134 Palermo, Italy
\and 
Consorzio COMETA, Via S. Sofia 64, 95123 Catania, Italy
\and
Bogolyubov Institute for Theoretical Physics, Metrologichna str. 14-b Kiev 03680, Ukraine
\and
Astronomical Observatory, Kiev National Taras Shevchenko University, Kiev 04053, Ukraine
\and
European Space Operations Centre, ESA, Robert-Bosch-Str. 5, D - 64293 Darmstadt, Germany 
\and
Institute for Applied Problems in Mechanics and Mathematics, Naukova St.\ 3-b, Lviv 79060, Ukraine
\and
Astronomical Observatory, National University, Kyryla and Methodia St.\ 8, Lviv 79008, Ukraine
\and
Instituto de Astronom\'{\i}a y F\'{\i}sica del Espacio (IAFE), CC 67, Suc. 28, 1428 Buenos Aires, Argentina
}

\date{Received, accepted}

\authorrunning{M. Miceli et al.}
\titlerunning{SN 1006}

\abstract
{Efficient particle acceleration can modify the structure of supernova remnants. In this context we present the results of the combined analysis of the \emph{XMM-Newton} EPIC archive observations of SN~1006.}
{We aim at describing the spatial distribution of the physical and chemical properties of the X-ray emitting plasma at the shock front. We investigate the contribution of thermal and non-thermal emission to the X-ray spectrum at the rim of the remnant to study how the acceleration processes affect the X-ray emitting plasma.}
{We perform a spatially resolved spectral analysis on a set of regions covering the whole rim of the shell and we exploit the results of the spectral analysis to produce a count-rate image of the ``pure'' thermal emission of SN~1006 in the 0.5-0.8 keV energy band (subtracting the non-thermal contribution). This image significantly differs from the total image in the same band, specially near the bright limbs.}
{We find that thermal X-ray emission can be associated with the ejecta and we study the azimuthal variation of their physical and chemical properties finding anisotropies in temperature and chemical composition. Thanks to our thermal image we trace the position of the contact discontinuity over the whole shell and we compare it with that expected from 3-D MHD models of SNRs with unmodified shock.}
{We conclude that the shock is modified everywhere in the rim and that the aspect angle between the interstellar magnetic field and the line of sight is significantly lower than $90^\circ$.}


\maketitle

\section{Introduction}
\label{Introduction}

The diffusive Fermi acceleration process in Supernova Remnant (SNR) shocks can account for the observed spectrum of galactic cosmic rays at least up to the knee at $3\times10^{15}$ eV and even beyond (\citealt{bv07}, \citealt{be87}). The $ASCA$ detection of synchrotron X-ray emission in SN~1006 has shown that electrons can be accelerated up to very high energy at the shock front, thus providing a direct proof for the connection between SNRs and cosmic ray acceleration (\citealt{kpg95}).

The loss of energy ``deposited'' in the production of relativistic particles and their non-linear back-reaction on the background plasma are predicted to strongly affect the structure of the remnant, by increasing the shock compression ratio well over the Rankine-Hugoniot limit and decreasing both the post-shock temperature and the geometric extension of the region between the main shock front and the contact discontinuity (\citealt{be99}, \citealt{deb00}, \citealt{bla02}, and references therein). All these effects modify the spectral properties of the thermal emission. 
Therefore, both thermal and non-thermal emission can provide important information on the acceleration processes and can be useful to verify the expected effects of the shock modification and to study the open questions still under debate, for example: what are the factors that influence the fraction of particles injected in the acceleration process? What is their maximum energy? What is the relationship between the upstream/downstream magnetic fields and the acceleration processes? 

To this end SN~1006 is a privileged target. Its rather simple morphology, the exact knowledge of its age, and its evolution in a low density uniform medium (because of its high galactic latitude) makes it an ideal laboratory for these studies. The bilateral morphology of SN~1006 shows two opposed radio and X-ray bright limbs characterized by non-thermal emission separated by a region with low surface brightness. The synchrotron limbs reveal narrow filaments both in radio and in X-rays with the X-ray emission presenting more pronounced inhomogeneities in surface brightness (see, for example, \citealt{rbd04}). As pointed out by \citet{rbd04}, the very low surface brightness above 2 keV in the interior strongly suggests that the ambient magnetic field is parallel to the shock velocity in the bright limbs (polar-cap scenario), in the framework of the quasi-parallel model for the injection (i. e.  the injection efficiency of the accelerated particles increases with decreasing the obliquity angle, $\psi$, between the magnetic field and the shock velocity). \citet{rbd04} studied the azimuthal variations of the spectral properties of the non-thermal emission over the whole shell and its radial profile from the center to the north-eastern bright limb. They found that the roll-off frequency of the synchrotron emission is much larger in the bright non-thermal limbs than in the center and elsewhere in the rim. These results are difficult to explain in the polar cap scenario, but are quite unexpected also in the ``equatorial belt" scenario (i. e. the ambient magnetic field, $B_0$, is oriented from South-East to North-West and the injection efficiency is larger when $\psi\sim 90^\circ$), which predicts variations of a factor $\sim2$, that is much less than the observed factor $>10$. Recently, \citet{pdc08} have shown that, by assuming a uniform ambient magnetic field in the polar-cap scenario, SN~1006 should be a centrally filled SNR in radio to reproduce the observed azimuthal radio profiles and this is contrary to what is observed. The determination of the geometry of the magnetic field in SN~1006 and its effects on the acceleration process are therefore an open problem and the study of thermal emission may provide useful indications, since it can be strongly affected by shock acceleration processes.

The X-ray thermal emission is, in comparison to the non-thermal, more uniformly distributed over the whole remnant (\citealt{rbd04}). Despite the importance of thermal X-rays for the study of shock acceleration processes, the physical origin of the thermal emission is still uncertain and the spatial distribution of its properties has not been studied yet.

Several studies have shown that thermal X-rays in SN~1006 cannot be uniquely associated with solar-abundance shocked plasma (e. g. \citealt{drb04}). X-ray emitting ejecta have been observed in the north-western and south-eastern rims of the remnant (\citealt{abd07}) and even Fe K lines emission has been recently detected in the interior of the shell, where iron overabundance has been revealed (\citealt{ykk08}). The presence of significant X-ray emission from the shocked interstellar medium (ISM) and the value of its temperature are indeed still controversial. \citet{abd07} modelled the X-ray emission of their spectral regions with one non-thermal component plus two thermal components (ejecta for the soft emission and ISM for the hard emission). However, they found that the shocked ISM component (with $kT_{ISM}\sim 1.5-2$ keV) is almost not needed from a statistical point of view, since the quality of the fits does not change by associating the thermal emission only with the ejecta. \citet{ykk08}, instead, used three thermal components to model the \emph{SUZAKU} thermal emission of very large regions of SN~1006, by associating the soft component with the shocked ISM ($kT_{ISM}\sim0.5$ keV), and the hot components with the ejecta. Nevertheless they cannot exclude that the O line complexes, dominating the soft component, originate in the ejecta.

A first approach to study the shock modification effects through the X-ray thermal emission was pursued by \citet{chr08} (hereafter C08). They focussed on the south-eastern quadrant of the shell and they assumed the emission in the $0.5-0.8$ keV band to be completely associated with the ejecta. They measured the ratio, $r_{BW/CD}$, of radii between the blast wave shock front (BW) and the contact discontinuity (CD) in several points, finding very low values with a minimum $r_{BW/CD}\sim1.0$ near the non-thermal limbs. The azimuthal trend of $r_{BW/CD}$ seems to be in agreement with the polar cap scenario, although the values of $r_{BW/CD}$ are too low and not consistent with that model.

Since a spatially resolved study of the thermal emission at the rim of SN~1006 has not been performed yet, here we present the analysis of archive \emph{XMM-Newton} observations of SN~1006. We focus on the rim of the remnant to study the azimuthal variations of thermal and non-thermal X-ray emission immediately behind the main shock front. We aim at obtaining detailed information on the nature of the thermal emission and to study its link with the shock acceleration processes. We also use a $Chandra$ archive observation of the north-eastern limb of SN~1006 taking advantage of its larger spatial resolution to test our conclusions.
The paper is organized as follows: the data and the data analysis procedure are described in Sect. \ref{Data processing}, in Sect. \ref{The data analysis} we present the X-ray results in terms of spatially resolved spectral analysis and image analysis. Our results are discussed in Sect. \ref{Discussion} and our conclusions are summarized in Sect. \ref{Summary and conclusions}.

\section{Data processing}
\label{Data processing}

\begin{center}
\begin{table*}[htb!]
\begin{center}
\caption{Relevant information about the data.}
\begin{tabular}{lccccc} 
\hline\hline
OBS$\_$ID  &   Date     &  MOS-pn $t_{exp}$ (ks)$^{*}$ &  RA, J2000         &  DEC, J2000  \\ \hline
0111090101 & 2000-08-20 &        7-3                   &$15^{h}03^{m}50^{s}$& $-41^\circ47'00''$     \\
0111090601 & 2001-08-08 &        7-3                  &$15^{h}03^{m}30^{s}$& $-42^\circ01'00''$    \\
0077340101 & 2001-08-10 &       30-21                  &$15^{h}01^{m}51^{s}$& $-41^\circ49'00''$  \\
0077340201 & 2001-08-10 &       24-19                  &$15^{h}04^{m}07^{s}$& $-41^\circ52'32''$  \\
0143980201 & 2003-08-14 &       16-11                   &$15^{h}03^{m}30^{s}$& $-41^\circ48'12''$   \\
0202590101 & 2004-02-10 &       26-17              &$15^{h}02^{m}35^{s}$& $-42^\circ04'37''$     \\
0306660101 & 2005-08-21 &       11-4                  &$15^{h}03^{m}35^{s}$& $-42^\circ04'20''$ \\
\hline\hline
\multicolumn{5}{l}{\footnotesize{* Unscreened/Screened exposure time.}} \\
\label{tab:data}
\end{tabular}
\end{center}
\end{table*}
\end{center}

We consider all the \emph{XMM-Newton} EPIC archive observations of SN~1006 available, but we do not present here the observation 0111090601 since it is strongly affected by proton flares and its cleaned exposure time is quite low ($\la 1$ ks). All the observations that we analyze were performed with the Medium filter by using the Full Frame Mode for the MOS cameras (\citealt{taa01}) and the Extended Full Frame Mode for the pn camera (\citealt{sbd01}).
The relevant information about the \emph{XMM-Newton} observations presented here are summarized in Table \ref{tab:data}.
We process the data by using the Science Analysis System (SAS V7.1). Light curves, images, and spectra, were created by selecting events with PATTERN$\le$12 for the MOS cameras, PATTERN$\le$4 for the pn camera, and FLAG=0 for both. To reduce the contamination by soft proton flares, we have screened the original event files using the sigma-clipping algorithm suggested by \citet{sk07}. The screened exposure times are given in Table \ref{tab:data}. 

We produce the images of the whole remnant by superposing (we use the $EMOSAIC$ task) the MOS1, MOS2, and pn images of the pointings shown in Table \ref{tab:data}. We have taken into account the differences between MOS and pn effective areas and the mosaiced images are MOS-equivalent. All the images are background-subtracted, vignetting-corrected, and adaptively smoothed (with the task $ASMOOTH$). The exposure and vignetting corrections were performed by dividing the count images by the corresponding superposed exposure maps (obtained with the task $EEXPMAP$).

Spectral analysis was performed in the energy band $0.5-5$ keV using XSPEC. To check the validity and the robustness of our result we decide to adopt two different procedures for the spectral extraction and for the background subtraction. 

\noindent \textbf{1.} We use the event files generated with SAS. We generate the Ancillary Response Files with the SAS $ARFGEN$ task, and we use the $EVIGWEIGHT$ task (\citealt{ana01}) to correct vignetting effects. The contribution of the background is derived from the high statistics background event files discussed by \citet{cr07}. We subtract the background contribution from the same region positions on the CCD (i. e. in the detector coordinates), in order to take into account the inhomogeneous response of the instruments across the field of view. For each of the spectral regions presented in Sect. \ref{The data analysis}, we extract the spectrum from the corresponding event file with the largest exposure time. Spectra were rebinned to achieve a signal-to-noise ratio per bin $>5\sigma$ and the fittings were performed simultaneously on both MOS spectra and on the pn spectrum. The reported errors are at 90\% confidence.

\noindent \textbf{2.} We use the Extended Sources Analysis Software (XMM-ESAS V2.0) which allows us to subtract instrumental and cosmic backgrounds separately. XMM-ESAS models instrumental background from the ``first principles'', using filter-wheel closed data and data from the unexposed corners of archived observations. The cosmic background is then modeled in XSPEC after the instrumental background subtraction. 
The standard XMM-ESAS method of background subtraction has some difficulties, namely the large number of parameters and lack of suitable model for the Al K$_\alpha$ ($\sim 1.49$ keV) and Si K$_\alpha$ ($\sim 1.75$ keV) instrumental lines. To overcome these difficulties, we develop a procedure which is similar to the ``double background subtraction'' technique (\citealt{rp03}). In this procedure, for each observation we produce spectra and response files from the ``background'' region, which lies out of SN~1006. Then we model the ``background'' spectra in XSPEC, and keep these ``background'' parameters fixed when fitting ``source'' spectra afterwords 
 To improve our knowledge of local cosmic background parameters, we add to them the RASS background data\footnote{They were obtained using HEASARC background tool, \url{http://heasarc.nasa.gov/cgi-bin/Tools/xraybg/xraybg.pl}.}, as suggested in XMM-ESAS manual. The obtained best-fit parameters, describing cosmic background, remaining soft proton contribution, and instrumental lines, are then fixed and used to model the parameters of "source" regions, after proper rescaling. To prepare the EPIC MOS event lists we use the XMM-ESAS script MOS-FILTER. We obtain the MOS1 and MOS2 spectra and construct the corresponding background using the XMM-ESAS scripts MOS-SPECTRA and XMM-BACK, respectively. Finally, we group the spectra and the corresponding background to have at least 25 counts per bin.

% The number of fitting parameters significantly increases, hence it is hard to find a true minimum of $\chi^{2}$. The quantitative analysis of the $1.3{-}1.8$ keV range is also not possible because of the presence of two strong unmodeled instrumental lines: 

We here present the results obtained with method 1, since it allows us to use also the pn data. We verified that the results obtained are perfectly consistent with that produced with method 2.

We also use a \emph{Chandra} archive observation of SN~1006 to perform the test described in Sect. \ref{Thermal emission of SN 1006}. In particular, we use the observation of the NE limb of SN1006 with ID 732, taken on 7 Jul 2000, which has an exposure time of 69 ks.
 
\section{The data analysis}
\label{The data analysis}
 
\subsection{Spatially resolved spectral analysis}
\label{spatially resolved spectral analysis}

\begin{figure}[htb!]
 \centerline{\hbox{     
     \psfig{figure=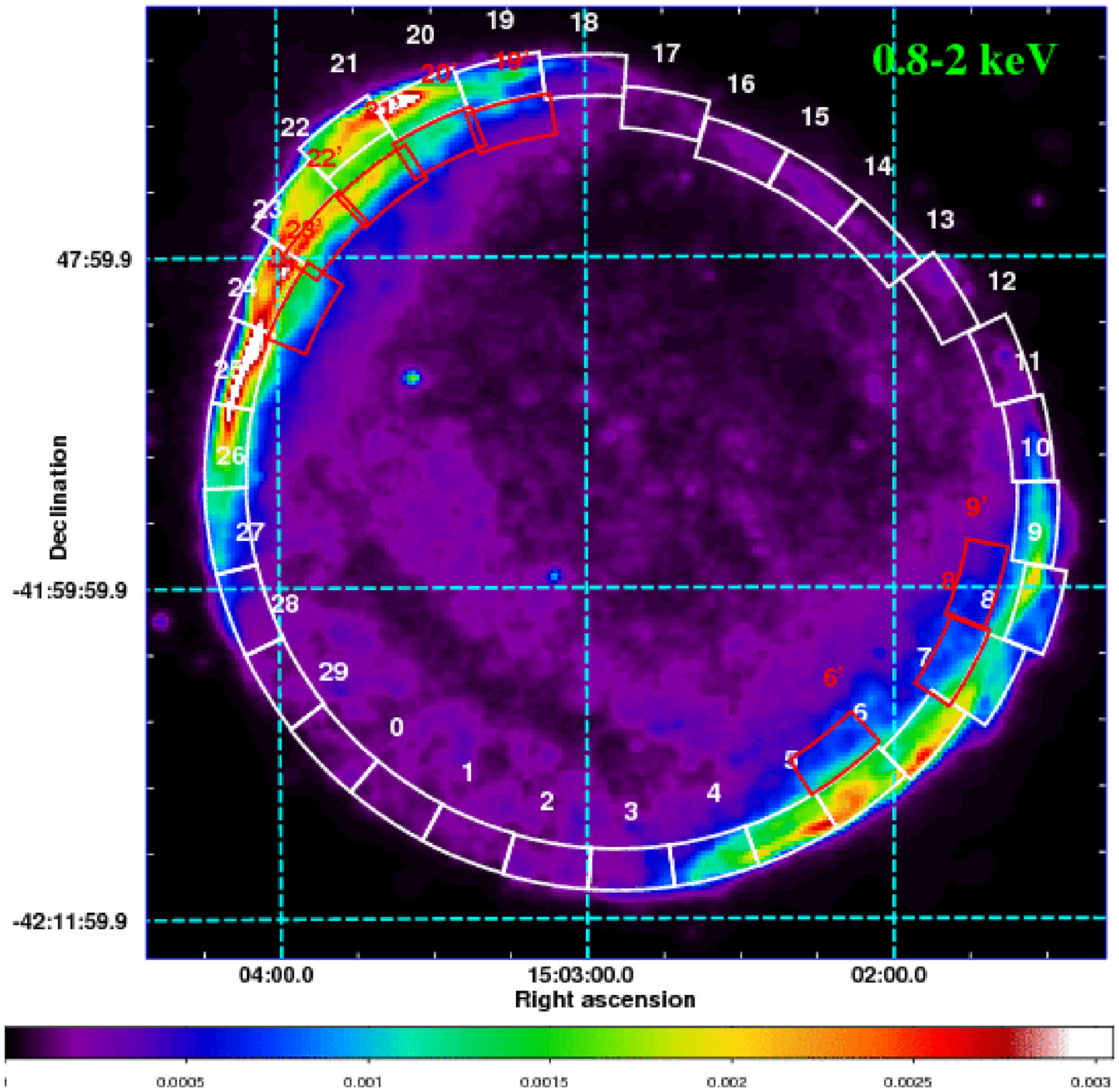,width=7cm}}}
      \centerline{\hbox{     
     \psfig{figure=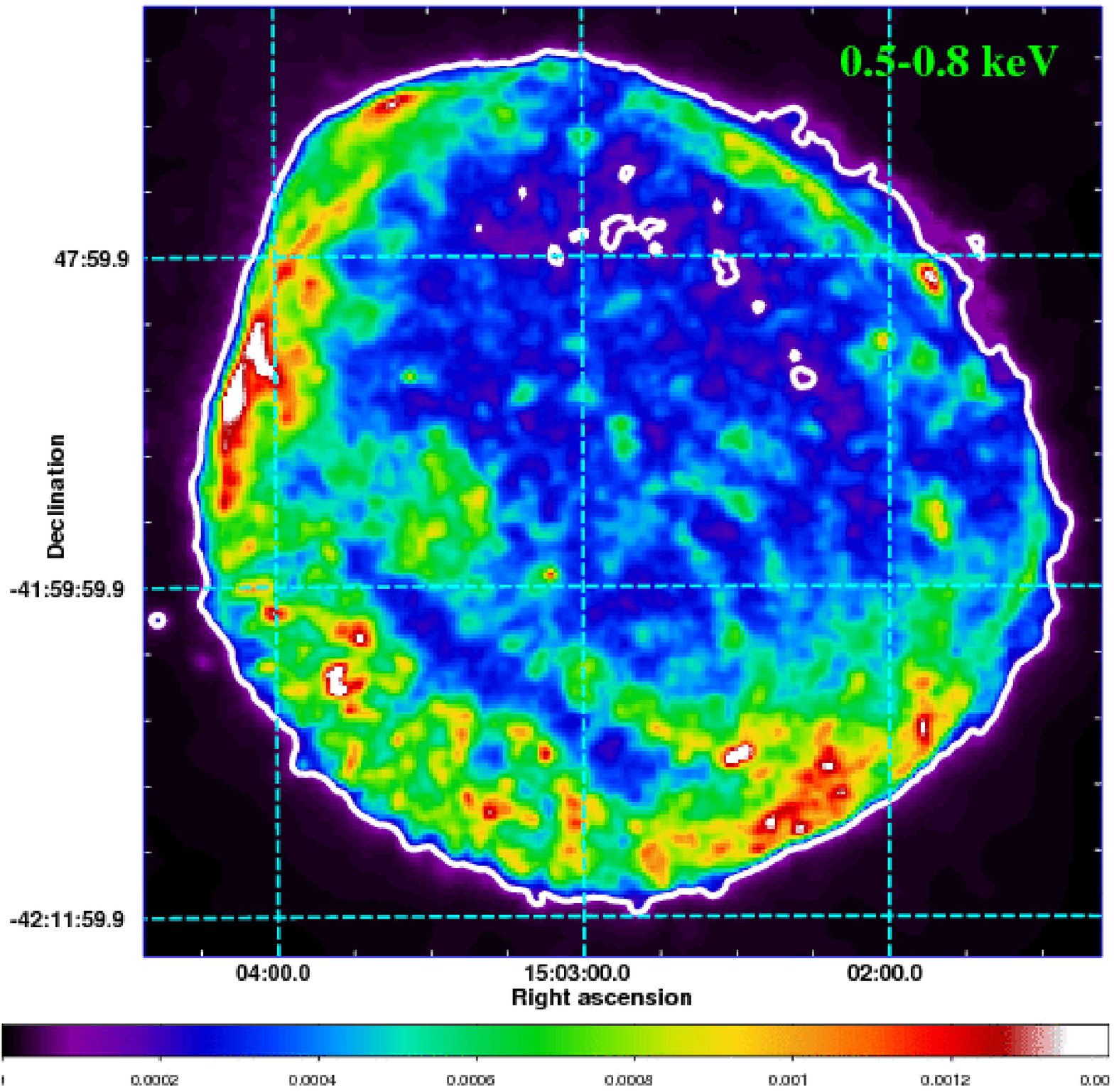,width=7cm}}}  
        \centerline{\hbox{     
     \psfig{figure=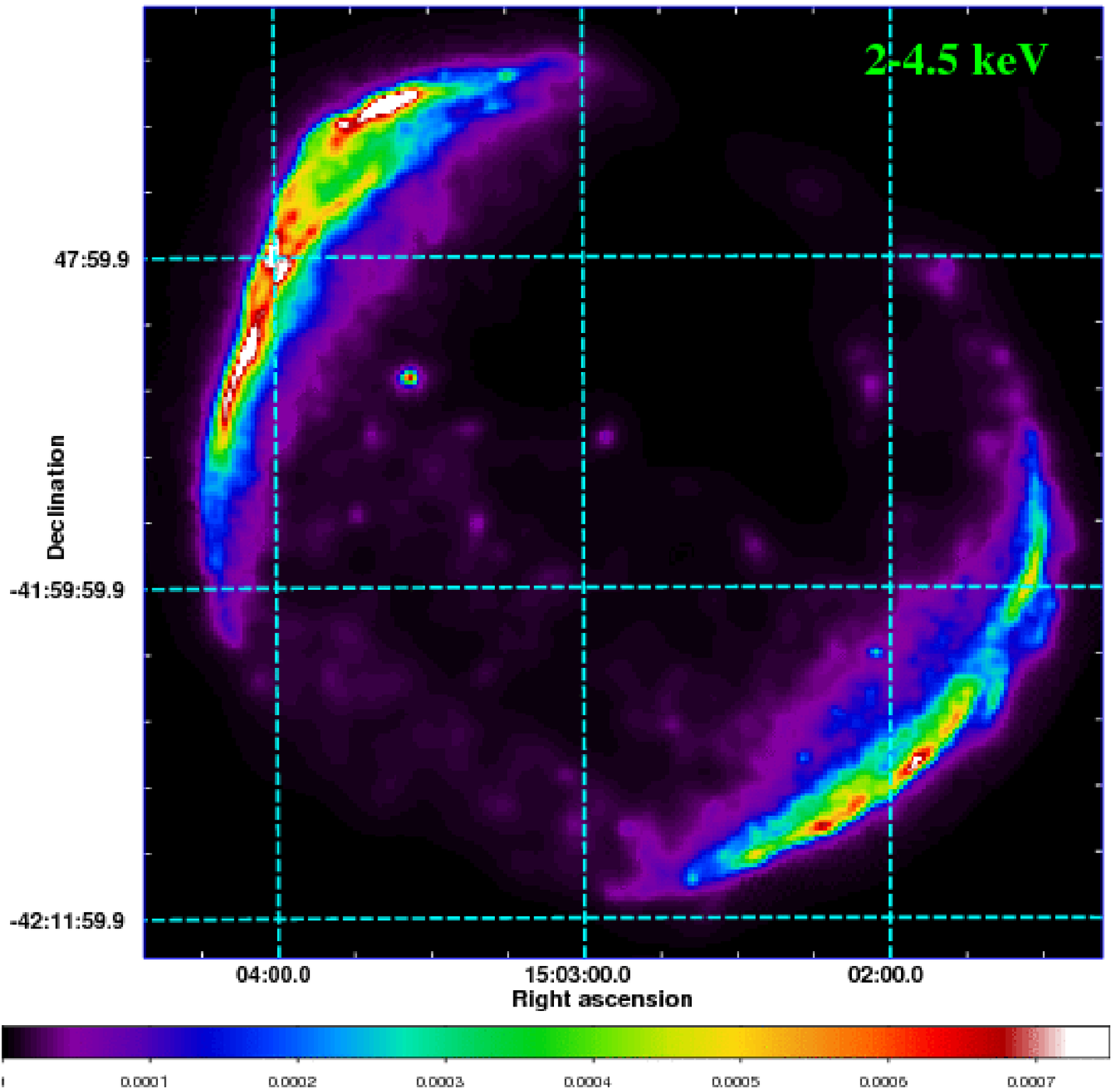,width=7cm}   
 }}
\caption{\emph{Upper panel}: mosaiced count-rate images (MOS-equivalent counts per second per bin) of SN~1006 in the $0.8-2$ keV band. The bin size is $8''$ and the image is adaptively smoothed to a signal-to-noise ratio 10. The 30 regions selected for the spectral analysis of the rim (in white) and the 8 regions selected for the study of the ejecta emission measure (in red, see Sect. \ref{Thermal emission of SN 1006}) are superimposed. North is up and East is to the left. \emph{Central panel:} same as upper panel in the $0.5-0.8$ keV energy band. A contour level at $10\%$ of the maximum is superimposed. \emph{Lower panel}: same as upper panel in the $2-4.5$ keV energy band.}
\label{fig:softhard}
\end{figure}

\begin{figure}[htb!]
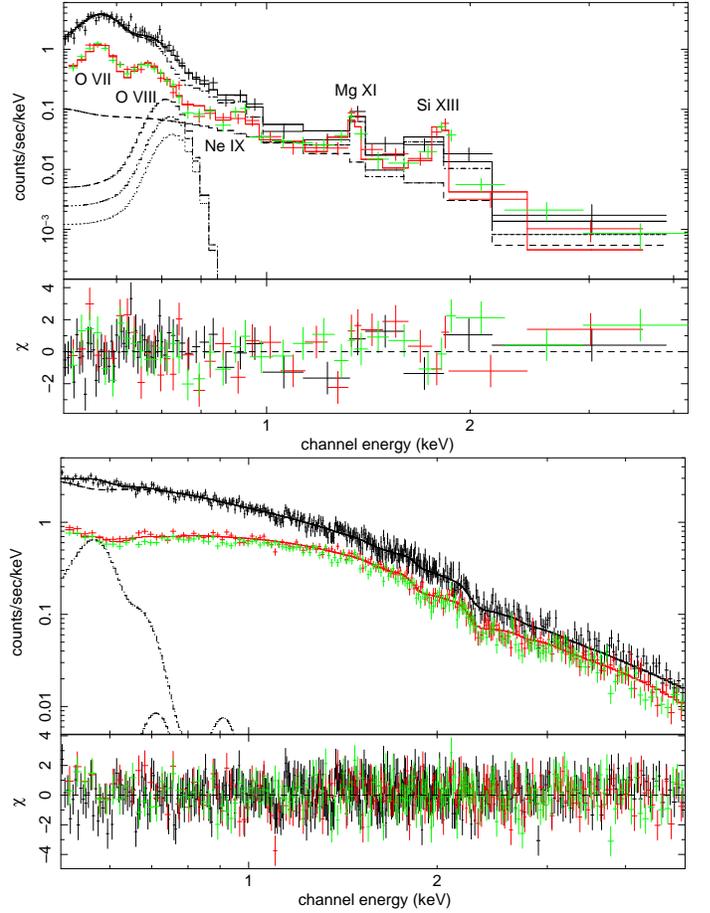

 \centerline{\hbox{     
     \psfig{figure=reg29add_lines2.ps,width=\columnwidth,angle=-90}}}
      \centerline{\hbox{     
     \psfig{figure=reg23add2.ps,width=\columnwidth,angle=-90}     
 }}
\caption{\emph{Upper panel}: pn spectrum (upper, in black) and MOS1,2 (lower, in red and green) spectra of region 29 of Fig. \ref{fig:softhard} with the corresponding best-fit model and residuals. The best-fit thermal$+$non-thermal model described in Sect. \ref{spatially resolved spectral analysis} is shown as continuous line, while individuals components are shown as dashed black lines for the pn spectrum only. Relevant emission lines are also indicated. \emph{Lower panel}: same as upper panel for region 23.}
\label{fig:specthnonth}
\end{figure}

\begin{figure}[htb!]
 \centerline{\hbox{     
     \psfig{figure=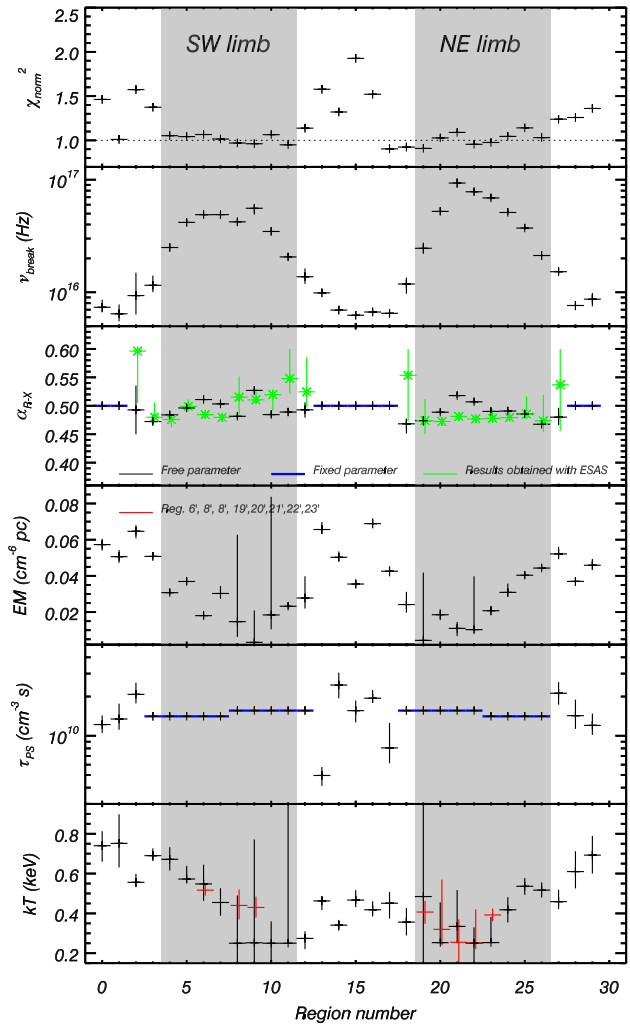,width=\columnwidth}}}
\caption{Best-fit parameters for the 30 regions shown in the upper panel of Fig. \ref{fig:softhard} (\emph{in black}) and for the 8 regions of the upper panel of the same figure (\emph{in red}). The model consists in a non-equilibrium of ionization isothermal component plus a synchrotron SRCUT model. The blue crosses indicate where the parameters are fixed. The errors are at $90\%$ confidence level. The chemical abundances are tied to those of region NW, for regions $8-22$, and to those of region SE, for regions $23-7$ (see Table \ref{tab:abund}). For comparison, the best-fit values of $\alpha$ obtained by using the ``ESAS approach" for the analysis of the background, as described in Sect. \ref{Data processing} are shown as \emph{green} stars.}
\label{fig:specres}
\end{figure}

\begin{figure}[htb!]
 \centerline{\hbox{     
     \psfig{figure=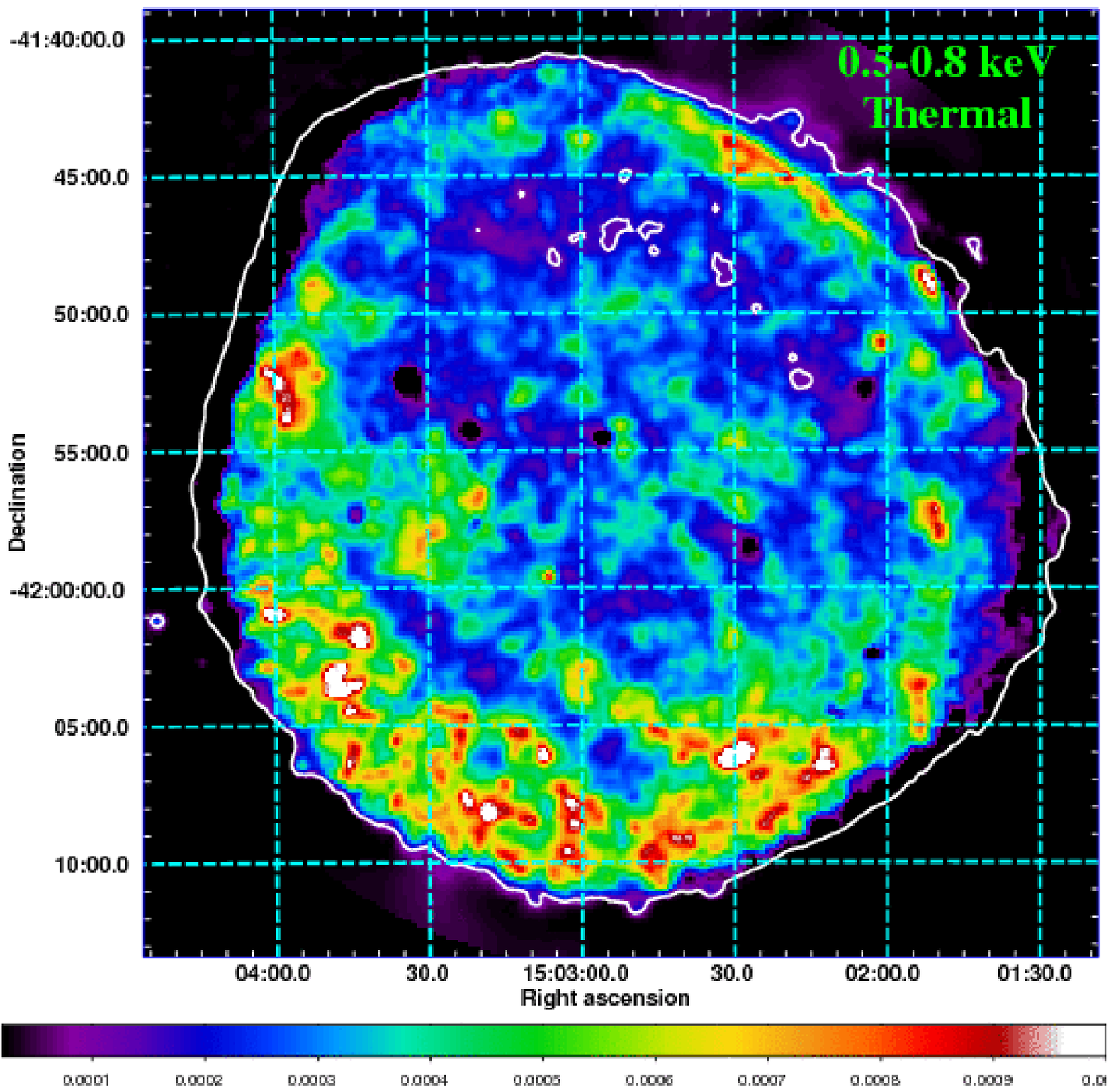,width=\columnwidth}}}
\caption{Mosaiced \emph{pure thermal image} (MOS-equivalent counts per second per bin) of SN~1006 in the $0.5-0.8$ keV band. The contribution of the non-thermal emission in the soft energy band has been subtracted. The bin size is $8''$  and the image is adaptively smoothed to a signal-to-noise ratio 10. The same contour level shown in the central panel of Fig. \ref{fig:softhard} is overimposed for comparison.}
\label{fig:thermal}
\end{figure}

Figure \ref{fig:softhard} shows the mosaiced EPIC images of SN~1006 in the $0.8-2$ keV band, $0.5-0.8$ keV band (``oxygen band'', central panel), and in the $2-4.5$ keV band (``hard band'', lower panel). Since we aim at studying the effects of the acceleration process, we select our spectral regions at the border of the shell. The set of 30 regions selected for the spatially resolved spectral analysis is shown (in white) in the upper panel of Fig. \ref{fig:softhard}. All the regions cover the same area in the plane of the sky and they extend in the radial direction for $1.5'$ (i. e. $\sim10\%$ of the SNR radius), corresponding to $\sim1$ pc at 2.2 kpc. Our approach is similar to that followed by \citet{rbd04}, although we here focus also on the thermal emission.

There are striking differences among the 30 spectra. Figure \ref{fig:specthnonth} shows the pn and MOS spectra extracted from region 29 (upper panel), where several emission line complexes are visible, and, for comparison, from region 23 (lower panel), located in the bright north-eastern non-thermal limb, where the emission is completely featureless. Despite these differences, we plan to describe all the spectra with a unified model, so as to explain the different spectral properties in terms of azimuthal variations of the best-fit parameters. We then use an isothermal model of optically thin plasma in non-equilibrium of ionization with a linear distribution of ionization timescale versus emission measure and with free abundances (VPSHOCK model in XSPEC, \citealt{blr01}) to describe the thermal component, plus a synchrotron emission from an electron power law with exponential cut-off  (SRCUT model in XPSEC, \citealt{rk99}) to model the non-thermal component. We verify that the quality of the fit does not improve by adding another thermal component. Moreover, an additional thermal component introduce too many free parameters (considering the available statistics) and this generates entanglements between the best-fit values, thus determining large errors and useless results. We therefore consider only one thermal component. Our model of the thermal emission is therefore different from the two-temperatures model adopted by \citet{abd07} for the north-eastern and south-western regions of the shell and it is simpler than the three-temperatures model used by \citet{ykk08}, where much larger and less uniform extraction regions are considered.

We constrain the normalization of the non-thermal component (i. e. its flux at 1 GHz) by extracting the radio flux from the same regions defined in the X-ray map. To this end we use a radio image obtained from interferometric VLA observations with the addition of single dish data. This procedure allowed us  to produce the highest angular resolution radio image of SN 1006, together with an accurate recovery of the flux density for all spatial structures. Details on the radio image can be found in \citet{pdc08} (see their Sect. 3 and Fig. 3). In the fittings of the X-ray spectra the radio flux at 1 GHz is fixed at its best value (derived from the radio map), but we verified that allowing it to vary within its uncertainties (it is determined with a $\sim20\%-50\%$ accuracy, depending on the spectral regions), the best-fit values and errors do not change significantly. The spectral index $\alpha$ is not fixed in our analysis, at variance with \citet{rbd04}, who fix $\alpha=0.6$.

We find relatively large residuals between data and spectral model in the $0.7-0.85$ keV band. The same problem has been observed in the $SUZAKU$ spectra of SN~1006 by \citet{ykk08} and it has been interpreted as a result of the presence of K-shell O~VII emission lines (transition series larger than $K_\gamma$) not included in the spectral code. To take into account the $K_\delta$, $K_\epsilon$, and $K_\zeta$ O~VII lines we follow the recipe by \citet{ykk08} and we add three narrow gaussians (a similar procedure has been adopted also by \citealt{abd07}). We fix the $N_H$ parameter to $7\times10^{20}$ cm$^{-2}$, in agreement with \citet{dgg02}.

The determination of the chemical abundances is complicated by the combined presence of thermal and non-thermal continuum. We focus only on the elements whose line complexes are visible in the spectra of the ``thermal'' regions (i. e. O, Ne, Mg, and Si\footnote{We fixed the other abundances to the corresponding photospheric solar values, except for the S abundance that is tied to that of Si.}, see upper panel of Fig. \ref{fig:specthnonth}), but we find that, even in thermal regions, the abundances are not well constrained. Although the relative ratios of O, Ne, Mg, and Si abundances are quite well determined in each thermal region, their absolute best-fit values are entangled with the normalization of the VPSHOCK component. In particular, for a given set of best-fit abundances, we obtain similar $\chi^2$ values by multiplying all the abundances by the same factor $k$ (even with $k=100$) and by reducing the normalization by the same factor. To avoid this degeneracy (that also hampers the determination of the emission measure), we consider two large regions: the first one is the sum of regions 28, 29, 0, and 1 (region SE); the second one is the sum of regions 13-16 (region NW). The larger statistics in these large regions allow us to accurately determine the chemical abundances, whose best-fit values are shown in Table \ref{tab:abund}. Notice that in the SE region the chemical composition of the X-ray emitting plasma is quite different from that in the NW region, where we find a lower Si$/$O ratio. In both cases, however, we observe significant overabundances. Moreover, the large Si$/$O ratios obtained in SE and NW regions are typical of type Ia SNe. We therefore conclude that the thermal component is associated with shocked ejecta. We assume regions $8-22$ to have the same abundances as region NW and the remaining 15 regions to have the same abundances as region SE.
\begin{center}
\begin{table}[htb!]
\begin{center}
\caption{Best-fit parameters.}
\begin{tabular}{lccccc} 
\hline\hline
            Element                 &     Region NW$^*$   &  Region SE$^{**}$   \\ \hline
          $O/O_\odot$               &    $5.0\pm0.1$      &    $4.4\pm0.3$      \\
          $Ne/Ne_\odot$             & $4.7^{+0.1}_{-0.2}$ & $1.5^{+0.2}_{-0.1}$ \\
       $Mg/Mg_\odot$                &   $12\pm2$          &    $15\pm1$         \\
       $Si/Si_\odot$                &  $29^{+3}_{-2}$     &       $50\pm3$        \\ \hline
        $kT$ (keV)                  &  $0.41\pm0.03$      &  $0.68^{+0.09}_{-0.04}$     \\
 $EM^{***}$ ($10^{17}$ cm$^{-5}$ )  &  $1.6\pm 0.1 $      &  $1.53_{-0.05}^{+0.12}$    \\
$\tau_{PS}$ ($10^{10}$ s cm$^{-3}$) &  $1.6\pm0.2$        &  $1.3^{+0.2}_{-0.3}$       \\   
   reduced  $\chi^2$ (d.~o.~f.)     &  $1.85$ $(426)$    &   $1.84$ $(234)$           \\   
\hline\hline
\multicolumn{5}{l}{\footnotesize{* Union of regions 13-16 of Fig. \ref{fig:softhard}.}} \\
\multicolumn{5}{l}{\footnotesize{** Union of regions 28, 29, 0, and 1 of Fig. \ref{fig:softhard}.}} \\
\multicolumn{5}{l}{\footnotesize{*** Emission measure per unit area.}} \\
\label{tab:abund}
\end{tabular}
\end{center}
\end{table}
\end{center}
 
The results of our spectral analysis are summarized in Fig. \ref{fig:specres}.

\subsubsection{Non-thermal emission}

As shown in Fig. \ref{fig:specres}, we find that the azimuthal profile of the synchrotron cut-off frequency $\nu_{break}$ presents a trend that is perfectly consistent with that derived by \citet{rbd04}, with the cut-off frequency increasing in the non-thermal limbs, although our  break frequencies are systematically lower than their ones (by a factor of $\sim10$). This discrepancy is due to the value of the radio spectral index $\alpha$. In fact, we here derive $\alpha\sim0.50\pm0.04$ from our spectral fittings in the non-thermal regions\footnote{In the thermal regions we therefore fix $\alpha=0.5$, as shown in Fig. \ref{fig:specres}.}, while \citet{rbd04} do not measure $\alpha$, and assume $\alpha=0.6$. The lower values of the spectral index naturally implies lower cut-off frequencies. Notice that, as shown by the green points in Fig. \ref{fig:specres}, we obtain $\alpha\sim0.5$ also by using the ``ESAS approach" for the analysis of the background, as described in Sect. \ref{Data processing}. Our best-fit value $\alpha\sim0.5$ and our cut-off frequencies are in agreement with the recent results obtained by \citet{ahs08}.

\subsubsection{Thermal emission}
Figure \ref{fig:specres} clearly shows that the emission measure per unit area, $EM$, of the thermal component significantly decreases in the bright non-thermal limbs (we obtain the same trend also analyzing the spectra with the ESAS method described in Sect. \ref{Data processing}). In particular, at North-East (regions 19-23) and at South-West (regions 6-10), $EM$ is a factor of $\ga4$ lower than at North-West or South-East and in regions 9 and 19 it is consistent with zero. This is quite surprising because one would expect $EM$ to be large in the non-thermal limbs, where the particle acceleration is supposed to be more efficient. In fact, \citet{deb00} have shown that, if the main shock is modified, the post-shock density is larger, even behind the reflected shock (and this is our case, since we are observing the shocked ejecta). We investigate in detail the reason and the physical origin of this result in the next section.

The temperature of the ejecta is not uniform and, in particular, in the north-western part of the rim (regions 13-18 in Fig. \ref{fig:specres}) the temperatures are clearly lower. Unfortunately, in the non-thermal limbs the determination of temperature is affected by large errors, because of the low $EM$. 

The ionization time-scale $\tau_{PS}$ has been determined only in thermal regions and is quite uniform, with a value around $10^{10}$ s cm$^{-3}$ (in the non-thermal regions we have fixed it to its mean value in the adjacent thermal regions). We observe a large deviation from the average value only in region 13, where $\tau_{PS}=5.0^{+0.8}_{-0.9}\times10^9$ s cm$^{-3}$. This is not surprising because region 13 contains the bright oxygen knot studied in detail by \citet{vlg03}, who found $\tau_{NEI}\sim2.35\pm0.07\times10^9$ s cm$^{-3}$. Indeed \citet{vlg03} used the NEI model (while we use the VPSHOCK model) and their value should be compared with $\tau_{PS}/2$ (see \citealt{blr01}), therefore our estimates are perfectly consistent with their ones. Notice also that the relative abundances found by \citet{vlg03} in the oxygen knot of ejecta are quite similar to those we derive for region SE, namely \citet{vlg03} find O:Ne:Mg:Si $\sim$ 0.18:0.05:0.2:1 in the oxygen knot and we find O:Ne:Mg:Si $\sim$ 0.1:0.03:0.3:1 in region SE (see Table \ref{tab:abund}). This result further confirms that the thermal component is associated with the ejecta.

\subsection{X-ray emitting ejecta in SN 1006}
\label{Thermal emission of SN 1006}

Since the thermal emission is completely dominated by the shocked ejecta, it is interesting to study their spatial distribution and to obtain an image of their X-ray emission. \citet{byk08} suggested that there are differences in the spatial distribution of thermal and non-thermal emission in the north-eastern region of SN~1006. \citet{chr08} assumed that (in the south-eastern quarter of the shell) the X-ray emission in the ``oxygen band'' ($0.5-0.8$ keV) is completely associated with the ejecta. Indeed, even in this soft band, there is a non-negligible flux associated with the non-thermal component, specially in the non-thermal limbs (see Fig. \ref{fig:specthnonth}). 
We use our results on the non-thermal component to estimate this contribution. 

The spectral properties of the SRCUT component are very robust and reliable and do not strongly depend on the model adopted for the thermal component, since the normalization of the synchrotron emission is derived from the radio data and the values of $\alpha$ and of the azimuthal profile of the roll-off frequency are in good agreement with those reported in literature. Using our result, we calculate that only in one third of the rim (regions $28-2$ and $18-22$) the contribution of the thermal emission in the oxygen band is larger than the $80\%$ of the total flux, while in fifteen regions ($4-11$ and $19-25$) it is $\le 50\%$. This means that the $0.5-0.8$ keV map of SN~1006 shown in the central panel of Fig. \ref{fig:softhard} cannot be used as a proxy of the ejecta.
\begin{figure}[htb!]
 \centerline{\hbox{     
     \psfig{figure=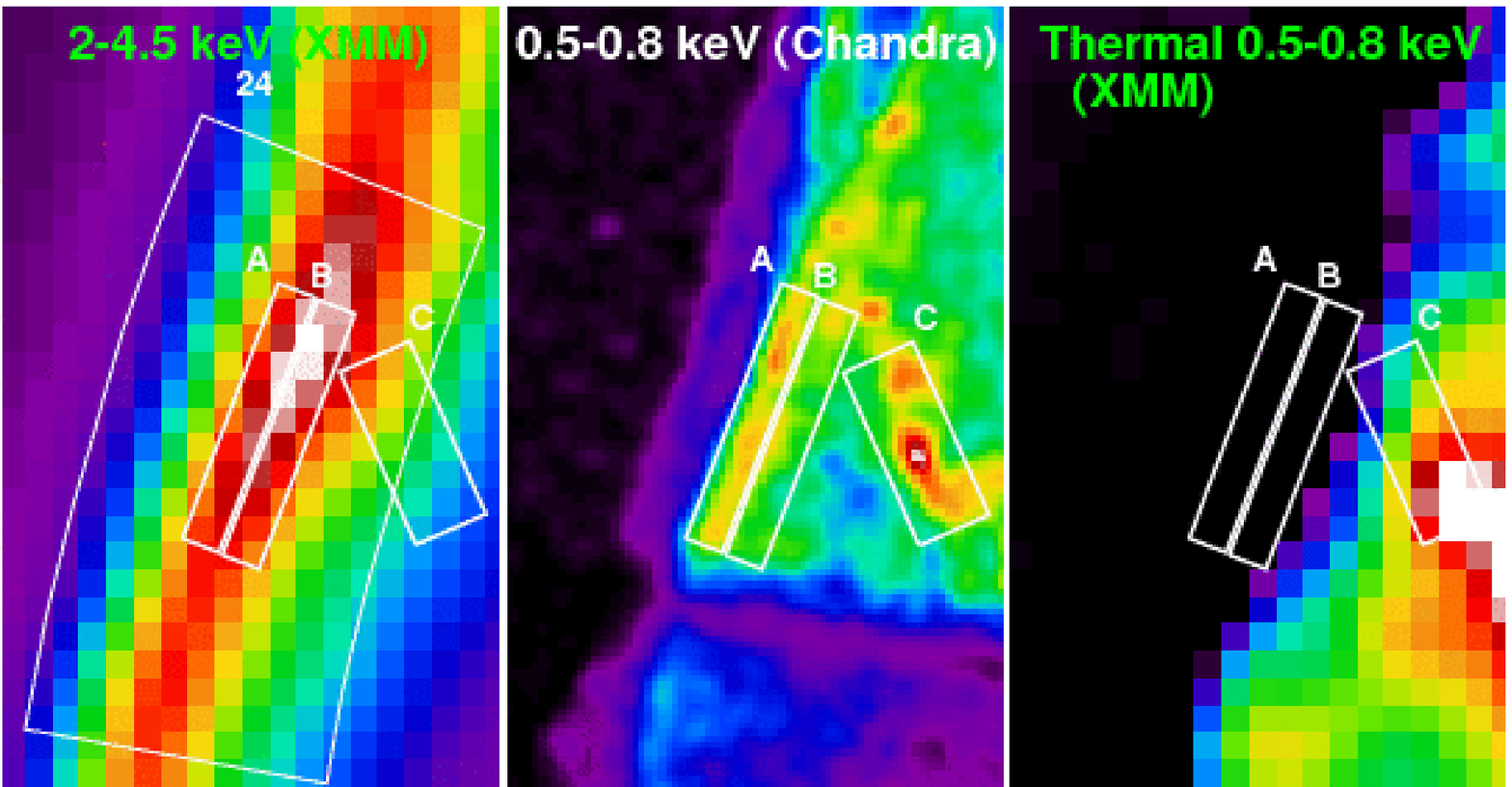,width=\columnwidth}}}
      \centerline{\hbox{     
     \psfig{figure=ChandraSpec.ps,width=\columnwidth,angle=-90}     
 }}
\caption{\emph{Upper panel}: Close-up view of region 24 of Fig. \ref{fig:softhard}. The \emph{XMM-Newton} count-rate images in the hard band is shown, together with the $Chandra$ ACIS-I image in the $0.5-0.8$ keV band compared with our \emph{XMM-Newton} thermal image in the same band (Fig. \ref{fig:thermal}). The three boxes indicate the regions where we extracted the $Chandra$ ACIS spectra. \emph{Lower panel}: $Chandra$ ACIS spectra extracted from the three regions shown in the upper panel}
\label{fig:chandra}
\end{figure}

The procedure that we use to produce the image of the ``pure" thermal emission in the oxygen band from our spectral results on the non-thermal emission is described in detail in Appendix A and is based on the assumption that the emission in the hard band ($2-4.5$ keV) is entirely non-thermal (for the validity of this assumption see Appendix A). Under this assumption, it is possible to use the spectral results to extrapolate the non-thermal emission in the oxygen band from the hard band image (in the appendix we show that the effects of the uncertainties in the best-fit parameters are negligible). Then, by subtracting the non-thermal contribution, we obtain the pure thermal image (i. e. the emission of the shocked ejecta) in the $0.5-0.8$ keV band. The map is shown in Fig. \ref{fig:thermal} and presents striking differences from the total image in the same band (central panel of Fig. \ref{fig:softhard}), specially in the non-thermal limbs, where the emission of the ejecta is negligible. 

In particular, in the spectral regions where the emission measure of the thermal component is low (e. g. regions 8, 9, 19, 21, 22, see Fig. \ref{fig:specres}), the thermal image indicates that the filling factor, $f$, of the ejecta is at its minimum. Therefore, it is the low volume filled by the X-ray emitting thermal plasma (in the spectral region) that causes the decreasing of the emission measure. In general, the comparison between Fig. \ref{fig:specres} and Fig. \ref{fig:thermal} reveals that there is a clear relationship between $f$ and $EM$: \emph{the variations of the emission measure of the thermal X-ray component can be associated with the different filling factors of the ejecta in the spectral regions}. 

To verify that this conclusion is correct and that our map of the soft emission of the ejecta is reliable, we perform two tests:

\noindent \textbf{1.} We consider 8 regions that, according to our thermal map, are expected to contain a larger filling factor of ejecta than regions $19-23$ and 6, 8, and 9. These new regions are indicated (in red) in the upper panel of Fig. \ref{fig:softhard}. Figure \ref{fig:specres} shows that the $EM$ in these new regions (indicated by the red crosses) is significantly larger than that of the corresponding regions on the rim, as predicted by our thermal image. Notice that the extension along the line of sight in these new regions is only a factor $\la10\%$ larger than in regions $19-23$ and 6, 8, and 9, therefore the much larger variations in $EM$ are, indeed, due to larger filling factor of the ejecta.

\noindent \textbf{2.} We exploit the large $Chandra$ spatial resolution to extract $Chandra$ ACIS-I spectra from the very narrow regions shown in Fig. \ref{fig:chandra}. Region A covers a thin non-thermal filament, region B is immediately behind, where the emission in the $0.5-0.8$ keV band is still large, but it is not associated with the ejecta, according to our thermal image, while region C covers a bright knot in our thermal image of the ejecta. The spectra extracted from these regions are shown in the lower panel of Fig. \ref{fig:chandra} and confirm the predictions of our thermal image: no O-line emission is visible in the spectrum of region B, that is very similar to that of the non-thermal region A, while in region C the oxygen line is clearly visible.

\section{Discussion}
\label{Discussion}

\begin{figure}[htb!]
      \centerline{\hbox{     
     \psfig{figure=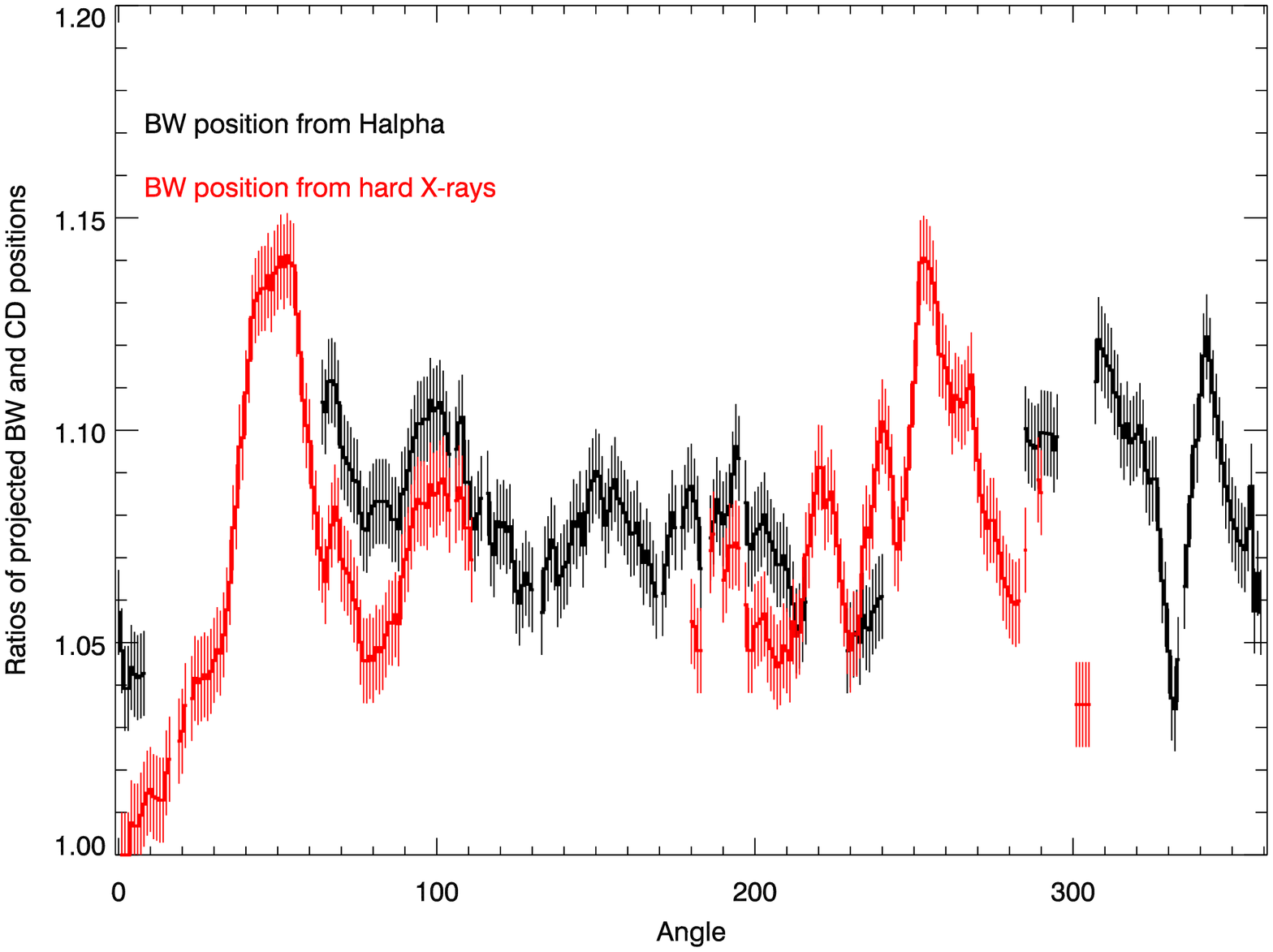,width=\columnwidth,angle=90}     
 }}
\caption{Azimuthal profile of the ratio of the position of the blast wave shock to the contact discontinuity with corresponding error bars. In order to reduce the fluctuations, for each angle we plot the moving average of the ratio, computed considering the ten adjacent points. The position of the contact discontinuity is derived from the pure thermal image in the oxygen band shown in Fig.\ref{fig:thermal}, while the position of the blast wave shock is derived from the $H_\alpha$ image of SN~1006 presented in \citet{wgl03} (black curve, missing points mark locations where $H_\alpha$ filaments are not well defined) and from the hard X-ray image shown in the lower panel of Fig. \ref{fig:softhard} (in red). The angle is measured counterclockwise from the North.}
\label{fig:bwcd}
\end{figure}

\begin{figure}[htb!]
 \centerline{\hbox{     
     \psfig{figure=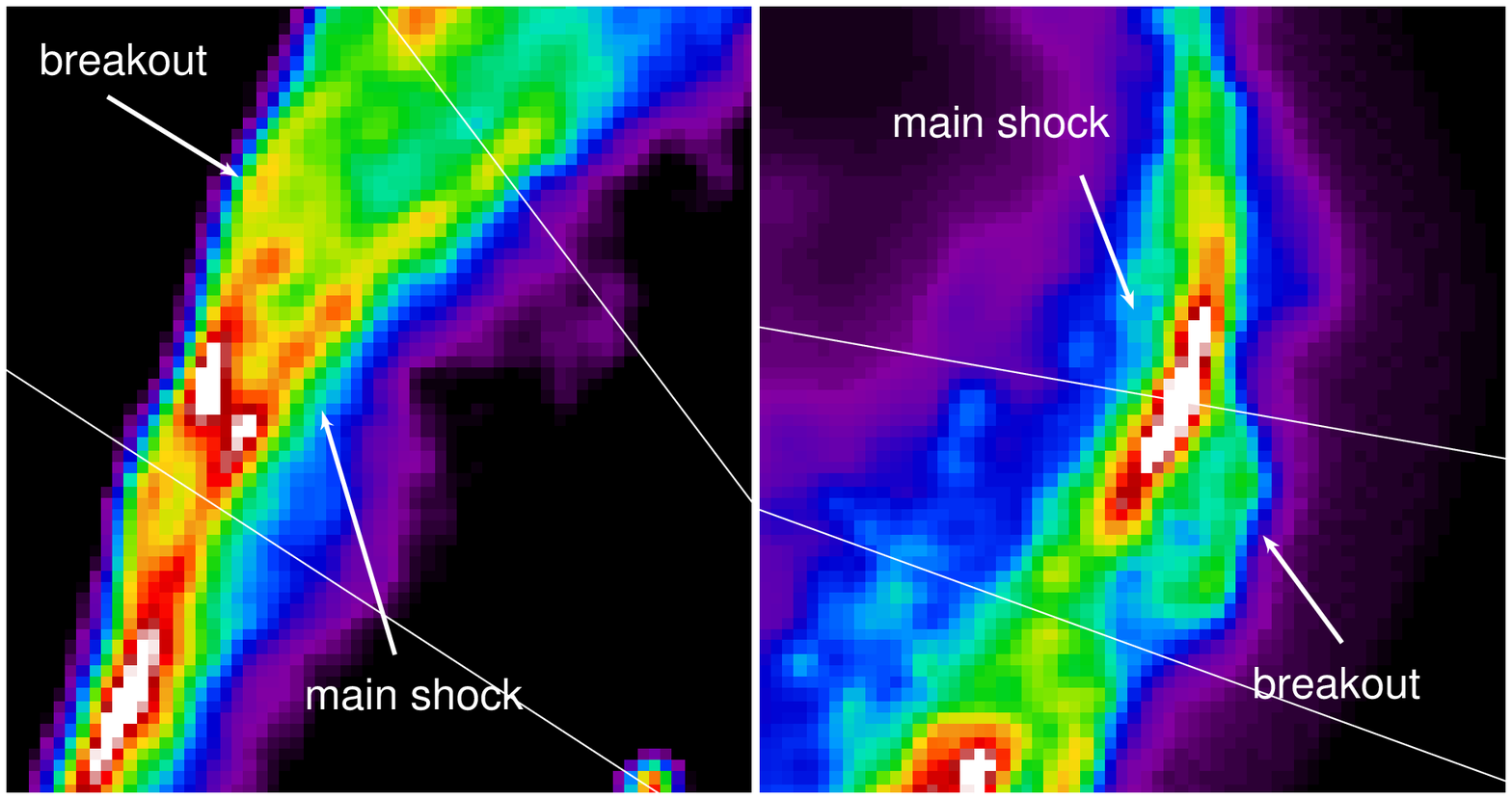,width=\columnwidth}}}
\caption{Close-up view of the $2-4.5$ keV band image of Fig. \ref{fig:softhard} showing two shock breakouts in the rim of SN~1006. We superimpose two lines indicating the azimuthal angles 40 and 60 (\emph{left panel}) and 250 and 260 (\emph{right panel})}
\label{fig:breakout}
\end{figure}
\subsection{Properties of the ejecta}
Our spatially resolved spectral analysis shows that thermal X-ray emission can be completely associated with the ejecta and the large Si$/$O ratios obtained in SE and NW regions are typical of type Ia SNe. Despite the overall spherical symmetry of the remnant, the distribution of the ejecta in the shell is not uniform. In particular, we detect inhomogeneities in the chemical composition of the ejecta, with the south-eastern part of the rim characterized by lower Si abundance and larger Ne abundance than the north-western part.
We also observe lower ejecta temperatures in the north-western part of the rim in agreement with the findings of \citet{abd07}. The interpretation of this result is not straightforward. As proposed by \citet{abd07}, in the North-West the remnant is interacting with a dense medium, originating a very bright H$_\alpha$ filament (\citealt{wgl03}), which is spatially correlated with the X-ray thermal emission. The physical conditions are then different from that in the South-East, where the expansion of the shell occurs in a more uniform medium. It is possible that the different thermal conditions between NW and SE are due to different heating mechanisms: in particular we can suppose the ejecta to be heated by the reverse shock at SE and by a reflected shock, generated by the interaction of the main shock front with the density enhancement at NW. According to this scenario the ISM density should be lower behind the dense cloud, thus causing an almost free expansion in that direction until the shock has reached the dense $H_\alpha$ cloud and generated the reflected shock. This may also explain the differences in the abundances of the ejecta, since it is possible that the reflected and reverse shock originated at different epochs (the interaction with the dense cloud is relatively recent) and have, therefore, interacted with different layers of the expanding ejecta. Nevertheless, according to this scenario, we would expect to find different ionization time-scales in the two regions, while we find that the ionization time-scales at NW are quite similar to those at SE (see Table \ref{tab:abund} and Fig. \ref{fig:specres}) and this seems to indicate that the inhomogeneities in the abundances are intrinsic (i. e. are the result of an anisotropic explosion) and not due to differences between the reverse and the reflected shock.

We can estimate the time elapsed after the shock impact, $t$, and the ejecta density $n_{ej}$. We determine $t$ from our best-fit values of $EM$ and $\tau_{PS}$ in the large NW and SE regions (Table \ref{tab:abund}). To this end we first derive the post-shock density of the ejecta from their emission measure per unit area. We first consider the NW and SE regions and, by assuming that the extension along the line of sight is $L\sim 1.85\times10^{19}$ cm (i. e. the length of the chord intercepted by the shell along the line of sight at the center of these spectral regions), we derive the density $n_{ej}\sim0.1/\sqrt{f}$ cm$^{-3}$ (where $f$ is the filling factor of the ejecta) both at NW and SE. We perform the same estimate in each of the thermal regions (regions $0-2$, $13-17$, and $27-29$) to check for the presence of azimuthal variations of the density, but we do not find any evidence for it. We then derive the time elapsed after the shock impact in the NW and SE regions, by using the relationship: $t\sim \tau_{PS}/n_{ej}\sim 2500\pm500~\sqrt{f}$ yr. Our value of $t$ is reasonable, considering that $f<1$, because of the clumpy nature of the ejecta.

\subsection{The distance between contact discontinuity and blast wave shock}
The image of the pure thermal emission of the ejecta, shown in Fig. \ref{fig:thermal}, allows us to trace the position of the contact discontinuity, CD, over the whole remnant. The distance of the CD from the blast wave position, BW, can be a powerful tracer for the presence of efficient particle acceleration at the blast wave shock. \citet{deb00} have shown that by increasing the injection efficiency $\eta$, the distance between BW and CD decreases. \citet{rbd04} have shown that the non-thermal X-ray emission of SN~1006 is consistent with the polar cap scenario and this implies that $\eta$ is larger in the non-thermal limbs, where, therefore, we expect the contact discontinuity to be closer to the shock front. C08 measured the distance between BW and CD in the southeastern quadrant of the shell, finding that the ratio BW/CD decreases towards the non-thermal limbs, as expected, but reaching too low values ($\sim1.0$). Nevertheless, to derive the position of the contact discontinuity, they assumed the emission in the $0.5-0.8$ keV band to be a tracer of the ejecta and we verified that this assumption is valid only in a small portion of the shell. We therefore adopt the same diagnostics used by C08, but we here use our pure thermal image in the oxygen band (Fig. \ref{fig:thermal}) to trace the contact discontinuity over the shell. 

We consider the radial profiles of the surface brightness in our thermal image for 360 different angles (the azimuthal angle increases counterclockwise from the North direction) and, for each angle, we identify the CD as the position of the relative maximum in brightness at the largest distance from the center\footnote{The positions of the blast wave shock and of the contact discontinuity are measured with respect to the same center as that chosen by C08}, having a surface brightness larger than a given threshold. We fix the threshold at $10\%$ of the flux of the brightest point in the pure thermal image. Since the pixel-size of the thermal map is $\sim 50\%$ of the EPIC point-spread function (FWHM), this procedure provides a quite accurate estimate of the CD position. The associated error is the addition in quadrature rule of the pixel size ($8''$) and of the sigma used to smooth the image (one pixel). 
The position of the main shock front in the thermal regions can be derived from the deep $H_\alpha$ image presented in \citet{wgl03}, corrected for the expansion of the remnant in the years elapsed from the optical observation to our observations (assuming an expansion speed of $4000$ km$/$s). As explained by C08, it is difficult to identify the optical rim by using a quantitative criterion (e. g. a threshold in the radial profile), so we determine the location of the BW, inspecting by eye the optical image. Nevertheless, it is not possible to use the $H_\alpha$ image to trace the position of the blast wave shock in the non-thermal limbs, since the optical emission is quite noisy and does not present sharp features. Therefore, in these regions, we use the $2-4.5$ keV image (lower panel of Fig. \ref{fig:softhard}) to trace the position of the blast wave shock, by adopting the same criterion as that used for the CD. 

Figure \ref{fig:bwcd} shows the azimuthal profile of the ratio between BW and CD, $r_{BW/CD}$, over the whole shell. In particular, to reduce the noise level, we plot for each angle the moving average of $r_{BW/CD}$ computed considering the ten adjacent points. By comparing our azimuthal profile of $r_{BW/CD}$ with that obtained in the southeastern quadrant by C08 we notice that, although a remarkable agreement in the thermal regions is present, our $r_{BW/CD}$ does not drop to very low values near the non-thermal limbs (at odds with C08 who finds $r_{BW/CD}\sim 1.0$). This is because the pure thermal image in the oxygen band differs significantly from the total $0.5-0.8$ keV image near the non-thermal limbs. The value of $r_{BW/CD}$ is quite uniform and does not present relevant azimuthal variations except for a couple of regions (between $\sim40$ and $\sim60$ degrees and between $\sim250$ and $\sim 260$ degrees) where it reaches two maxima at $\sim 1.15$. We verify that in these two regions remarkable shock breakouts are present, as shown in Fig. \ref{fig:breakout}. We interpret these breakouts as the result of the propagation of the main shock in a locally rarefied medium. If the shock has recently encountered a low density inhomogeneity its speed is larger with respect to the adjacent regions, while the position of the reverse shock and of the contact discontinuity are not immediately affected. This explains the locally large values of $r_{BW/CD}$. 
Elsewhere $r_{BW/CD}$ is limited in the range $r_{BW/CD}\sim1.05-1.12$.
\begin{center}
\begin{table}[htb!]
\begin{center}
\caption{Parameters of the 3-D MHD model}
\begin{tabular}{ll}
\hline
\hline
ISM density &  0.05 cm$^{-3}$ \\
Ambient magnetic field &  $30~\mu G$ \\
Explosion energy  &  $1.3 \times 10^{51}$ erg \\
Mass of the ejecta  &  $1.4~M_{\odot}$ \\
\hline
Shock speed after 1000 yr & 4750 km s$^{-1}$ \\
Shell radius after 1000 yr & 8.65 pc \\
\hline
\hline
%\multicolumn{2}{l}{\footnotesize{* Union of regions 13-16 of Fig. \ref{fig:softhard}.}} \\
\label{tab:mhd}
\end{tabular}
\end{center}
\end{table}
\end{center}

\begin{figure}[htb!]
 \centerline{\hbox{     
    \psfig{figure=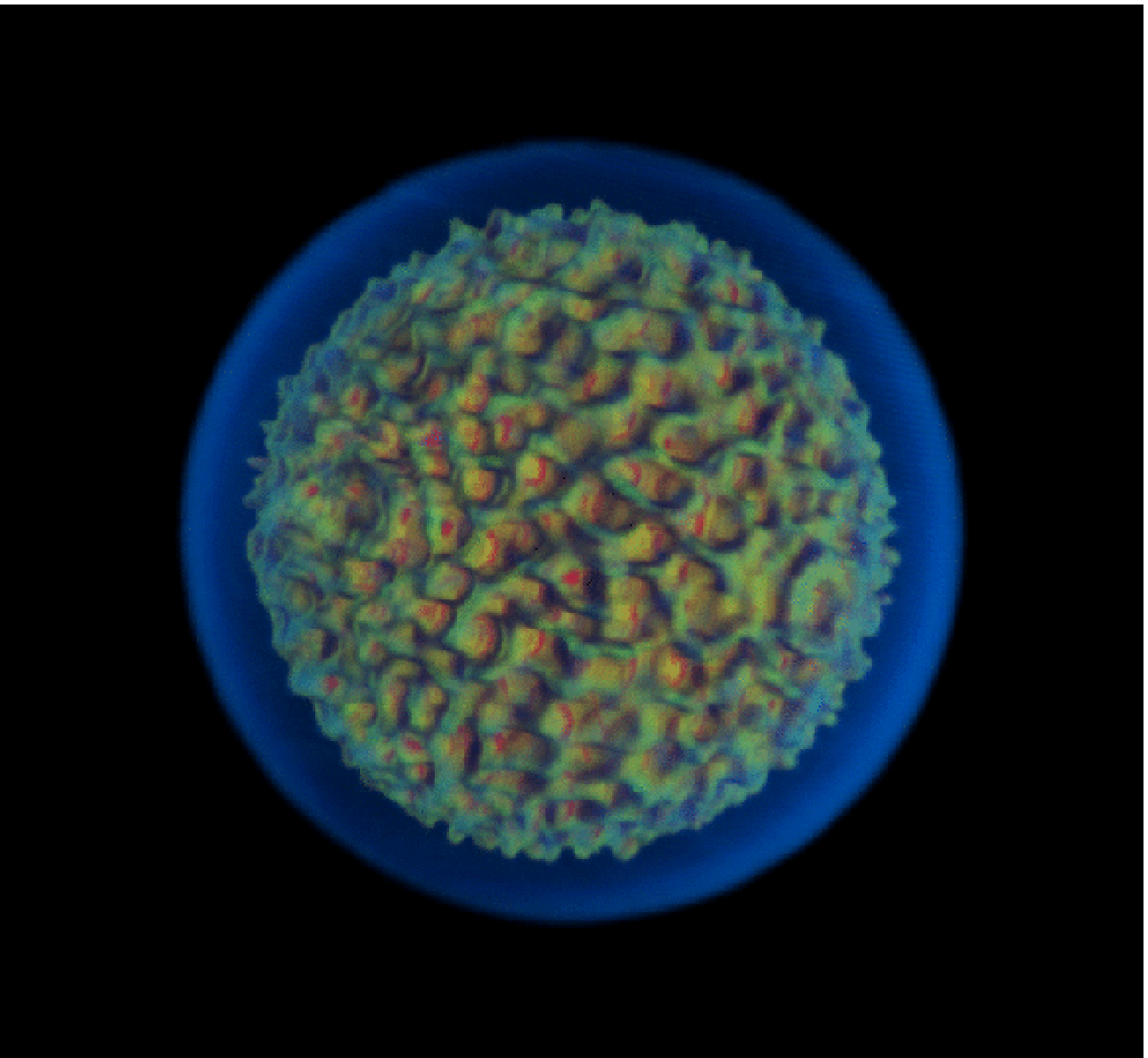,width=\columnwidth}}}
\caption{3-D rendering of an MHD simulation describing the expansion of SN1006 through a magnetized ISM at t=1000 yrs (the parameters of the model are shown in Table \ref{tab:mhd}).
The ejecta material is tracked with a ``solid" surface, the shocked ISM in semi-transparent blue. }
\label{fig:mhd}
\end{figure}

We verify that this value is not consistent with that expected for a non-modified Sedov shock. To this end we consider the 3-D magnetohydrodynamic model developed by \citet{obr07} to describe the expansion of a SNR through the magnetized ISM, adopting a set of parameters appropriate for SN~1006 (see Table \ref{tab:mhd}). The model includes no eventual magnetic field amplification, and no effects on shock dynamics due to back-reaction of accelerated cosmic rays. We explore different
configurations of the interstellar magnetic field by performing a set of numerical simulations using FLASH, an adaptive mesh refinement
multiphysics code (\citealt{for00}). In all the cases we find that Richtmyer-Meshkov (R-M) instability develops in the inner shell, as the forward and reverse shocks progress through the ISM and ejecta, respectively (e.g. \citealt{kdr99}). After 1000 yrs of evolution, the R-M instability determines a quite complex 3-D density and temperature structure of the contact discontinuity which extends a bit further with respect to the unidimensional case (where the instabilities are not described), leading to different values of the ratio $r_{BW/CD}$. The results of our simulation at t=1000 yrs are shown in Fig. \ref{fig:mhd}: the ``solid" surface is a tracer of the ejecta material, while the density of the shocked ISM is shown in semi-transparent blue.
In all the cases examined, we find that, according to our MHD model of a non-modified SNR shock, we expect $r_{BW/CD} > 1.16$, at $t=1000$ yrs. 
As shown in Fig. \ref{fig:bwcd}, we observe much lower values of $r_{BW/CD}$ in SN~1006. As shown by \citet{deb00} and \citet{edb04}, when efficient particle acceleration is present and a significant fraction of the shock energy is deposited in cosmic rays, the region between BW and CD becomes narrower, thus making $r_{BW/CD}$ lower. Since we observe much lower values of $r_{BW/CD}$ in the whole rim with respect to the non-modified case, we conclude that in SN~1006 the shock is modified everywhere.

If the ambient magnetic field were aligned in the SW-NE direction in the plane of the sky, with no component along the line of sight, we would expect lower values of $r_{BW/CD}$ in the non-thermal limbs, at odds with our findings. Nevertheless, if the aspect angle, $\phi$, between the interstellar magnetic field and the line of sight is significantly different from a right angle (i. e. the interstellar magnetic field is not in the plane of the sky and $\phi\la 70$ degrees), we will not observe any significant variations of $r_{BW/CD}$, since, in this case, we do not observe region where the acceleration process is more efficient edge-on. Therefore, our results suggests that the aspect angle in SN~1006 is significantly different from $90^\circ$. A similar conclusion has been obtained by \citet{pdc08} by comparing observed and synthesized radio maps of SN~1006.

\section{Summary and conclusions}
\label{Summary and conclusions}

\begin{enumerate}
\item The emission at the rim of the shell can be described by a mixture of thermal emission from plasma in non-equilibrium of ionization and non-thermal SRCUT emission. We compare two different approaches to X-ray spectra analysis, obtaining perfectly consistent results with the two methods.
\item We measure the radio-to-X-ray photon index of the non-thermal component, finding $\alpha\sim0.5$, and we confirm the azimuthal profile of the roll-off frequency found by \citet{rbd04}, revising the values of the frequencies in the light of the new value of $\alpha$.
\item Thermal emission can be associated with shocked ejecta. The temperature and the chemical composition of the ejecta are not uniform. In particular, we find lower temperatures and lower Si/O abundances at NW than at SE.
\item Our pure thermal image in the $0.5-0.8$ keV band shows that even in this soft band, the emission at the bright limbs is mainly associated with the non-thermal component.
\item We derive the azimuthal profile of the ratio $r_{BW/CD}$ between the position of the blast wave shock and the contact discontinuity in the whole shell. $r_{BW/CD}$ is fairly uniform and its mean value is lower than that expected from a non-modified shock, even taking into account hydrodynamic instabilities that make $r_{BW/CD}$ lower than that derived from simple 1-D hydrodynamic models. We conclude that the shock is modified everywhere and that the aspect angle is significantly different from $90^\circ$.
\end{enumerate}

\begin{acknowledgements}
We wish to thank F. Winkler for providing us the $H_\alpha$ image in digital format.
This work makes use of results produced by the PI2S2 Project managed by the Consorzio COMETA, a project co-funded by the Italian Ministry of University and Research (MIUR) within the Piano Operativo Nazionale ``Ricerca Scientifica, Sviluppo Tecnologico, Alta Formazione'' (PON 2000-2006). More information is available at http://www.pi2s2.it and http://www.consorzio-cometa.it. The software used in this work was in part developed by the DOE-supported ASC / Alliance Center for Astrophysical Thermonuclear Flashes at the University of Chicago. 
ESAS calculations were performed on computational resources of BITP computing cluster and VIRGO.UA Virtual Observatory. The work of D.I. was supported by the Program of Fundamental Research of the Physics and Astronomy Division of the National Academy of Sciences of Ukraine. DI and OP acknowledge the program "Kosmomicrofizyka" of the National Academy of Sciences of Ukraine. IT acknowledges the support from the INTAS YSF grant No. 06-1000014-6348. IT and DI thank ESAC (European Space Astronomy Centre) for hospitality where a part of this work
was done.
GD and GC are members of CONICET (Argentina). The project was partially funded by grants from  CONICET, UBA  and ANPCYT (Argentina)
\end{acknowledgements}

\bibliographystyle{aa}

\newpage

\appendix 

\section{Production of the pure thermal image}

We here describe the procedure we developed to produce the image of the thermal emission in the $0.5-0.8$ keV band (we call this image $TH$). The scheme of the procedure consists in three steps: A) we assume the emission in the $2-4.5$ keV band to be completely associated with the non-thermal component; B) we extrapolate the $2-4.5$ keV image to produce an image of the non-thermal component in the $0.5-0.8$ keV band, (we call this image $NONTH$) by using the results of our spatially resolved spectral analysis ; C)  we subtract $NONTH$ from the total image in the $0.5-0.8$ keV band, thus obtaining the thermal image. In the following we describe in detail these three steps.

\subsection{Step A (assumption)} 
\label{Step A}
We assume the thermal emission to be negligible in the $2-4.5$ keV band. Indeed the results of our spatially resolved spectral analysis show that this assumption is strictly valid in the non-thermal limbs (where the non-thermal flux is $\ga 99\%$ of the total), while it does not hold in the thermal regions (e.g. regions $28-2$), where about $50\%$ of the flux in the $2-4.5$ keV is associated with the thermal component. Nevertheless, we can show that this problem does not generate significant effects. In fact, even if we overestimate the contribution of the synchrotron emission in the thermal region (by a factor of a few) this contribution is still small in the $0.5-0.8$ keV band: for example we verified (after step B) that it is $\sim 10\%$ and $\sim6\%$ of the total flux in the oxygen band in region 0 and region 16, respectively. This means that, when we produce $TH$, the flux in the thermal regions is not significantly underestimated, while that in the non-thermal region is correct. In conclusion we can say that our assumption holds where it is more necessary.

\subsection{Step B}
\label{Step B}
\begin{figure}[htb!]
 \centerline{\hbox{     
     \psfig{figure=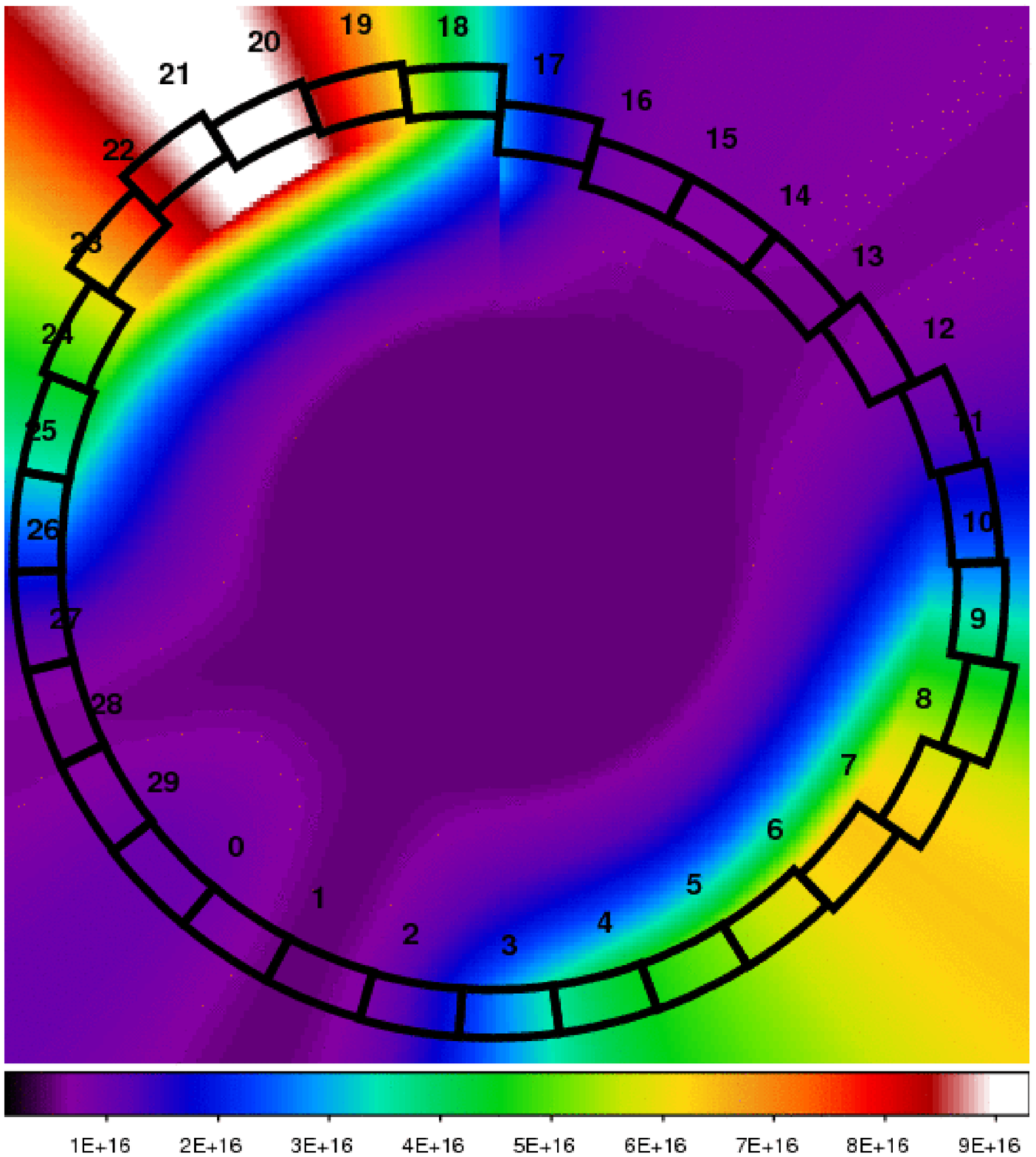,width=\columnwidth}}}
 \centerline{\hbox{     
     \psfig{figure=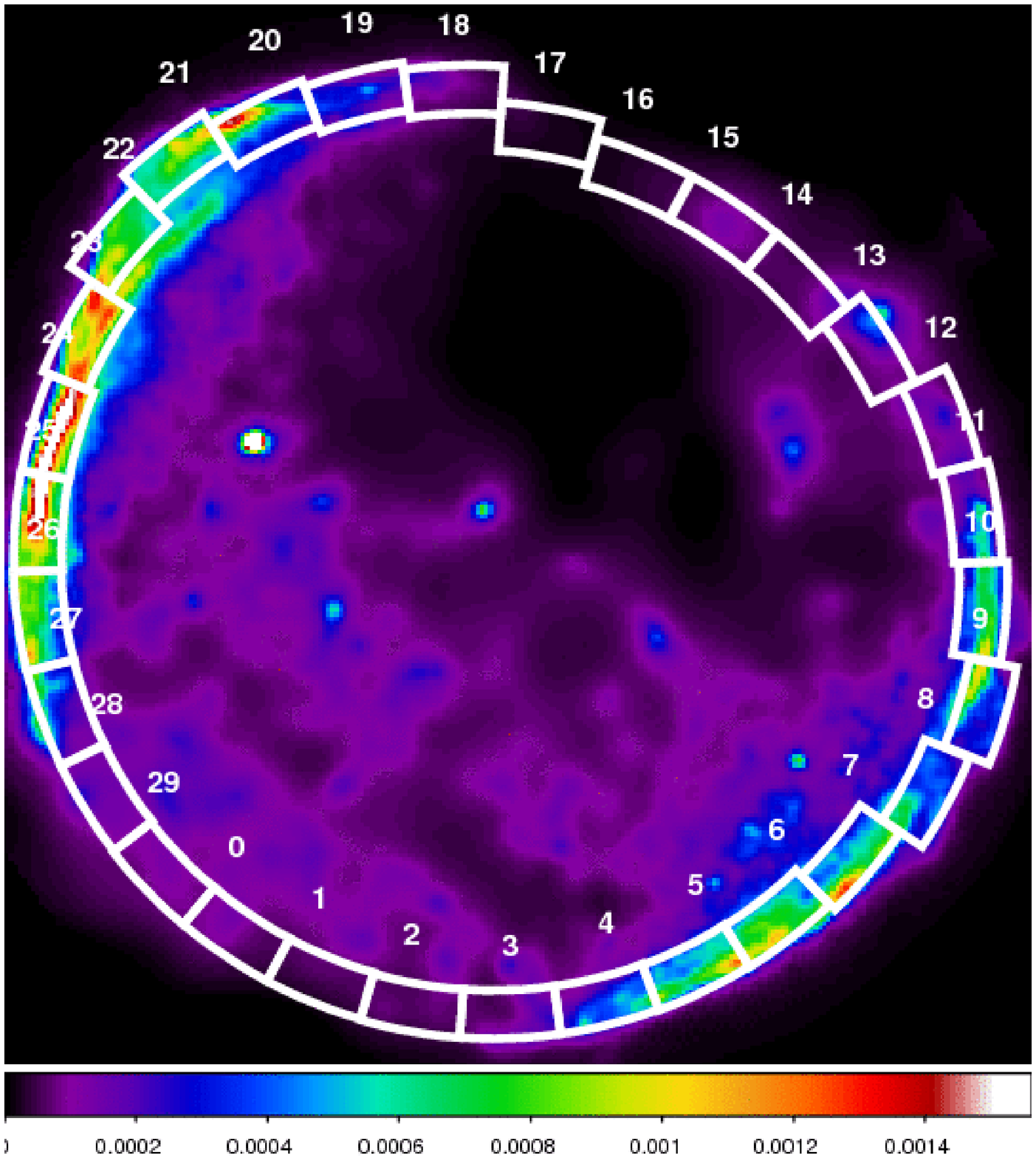,width=\columnwidth}}}  
\caption{\emph{Upper panel:} map of the cut-off frequency derived according to the procedure described in Sect. \ref{Step B}. The 30 regions selected for the spectral analysis of the rim are indicated in black. \emph{Lower panel:} image of the non-thermal emission in the $0.5-0.8$ keV band.}
\label{fig:cutoff}
\end{figure}

\begin{figure}[htb!]
 \centerline{\hbox{     
     \psfig{figure=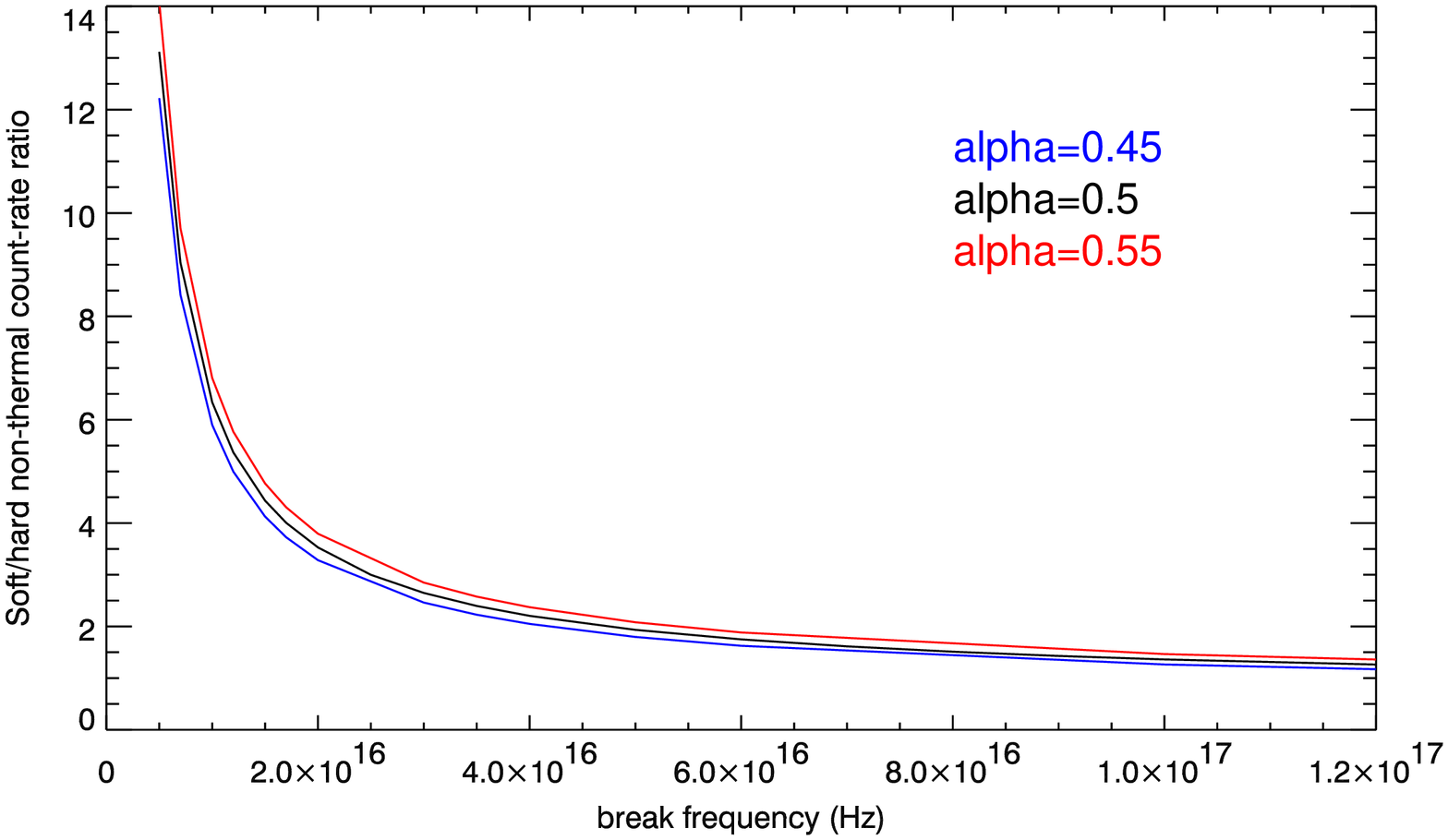,width=\columnwidth,angle=90}}}  
\caption{Ratio of the EPIC MOS count-rate in the $0.5-0.8$ keV and $2-4.5$ keV bands as a function of the break frequency, obtained by assuming an SRCUT model with spectral index $\alpha=0.45$ (blue curve), $\alpha=0.5$ (black curve), and $\alpha=0.55$ (red curve)}
\label{fig:ratio}
\end{figure}

From the $2-4.5$ keV image we can produce $NONTH$ by considering a given spectral shape of the synchrotron emission (i. e. a given value of $\alpha$ and a given value of the roll-off frequency $\nu_{break}$) in each pixel. Since we found that in the whole rim the photon index is fairly uniform (see Fig. \ref{fig:specres}), we can assume that in each point of the image in the hard band the spectrum has a photon index $\alpha = 0.5$. As for $\nu_{break}$, we found in Sect. \ref{spatially resolved spectral analysis} that it depends on the azimuthal angle $\theta$ at the rim. We then interpolate the values of $\nu_{break}$ shown in Fig. \ref{fig:specres} by using a Fourier series (up to the 6th order) and we assume that the azimuthal trend of $\nu_{break}(\theta)$ at the rim is given by this interpolating function at angular distances from the center $R>R_{int}(\theta)$, where $R_{int}(\theta)=11.7'$ for regions $13-17$ and $R_{int}(\theta)=13.5'$ elsewhere. As for the radial profile of $\nu_{break}(R)$ for $R<R_{int}(\theta)$, we follow \citet{rbd04} and we assume that: i) in the center ($R<7'$) the cut-off frequency is uniform and its value is equal to the minimum $\nu_{break}$ found in the rim and ii) for $7'<R<R_{int}(\theta)$, the cut-off frequency increases exponentially with $R$, going from its minimum value (at the center) to the corresponding value at the rim. In this way we produce the map of $\nu_{break}$ shown in Fig. \ref{fig:cutoff}.
Since we know the spectral shape of the synchrotron emission in each pixel of the $2-4.5$ keV image, we can produce the image of the non-thermal emission in the $0.5-0.8$ keV band. This image is shown in the lower panel of Fig. \ref{fig:cutoff}. We investigate how the uncertainties in the best-fit parameters influence $NONTH$. Figure \ref{fig:ratio} shows the ratio, $\rho$, of the non-thermal EPIC MOS count-rate in the $0.5-0.8$ keV and $2-4.5$ keV bands as a function of $\nu_{break}$, obtained with $\alpha = 0.5$ (black curve), $\alpha = 0.45$ (blue curve), and $\alpha = 0.55$ (red curve). This ratio provides the factor that allows us to produce, for each pixel, $NONTH$ from the $2-4.5$ keV map. Since in Figure \ref{fig:ratio} the red curve, the blue curve, and the black curve are very similar, we can conclude that the uncertainties in the determination of $\alpha$ do not significantly alter our results. As for the effects of the uncertainties in $\nu_{break}$, we focus only on the non-thermal limbs, since in thermal limbs, as explained in Sect. \ref{Step A}, the non-thermal count-rate in the $0.5-0.8$ keV band is negligible. Figure \ref{fig:ratio} shows that in the non-thermal limbs (where $\nu_{break} \ga 4\times10^{16}$ Hz and where the errors in $\nu_{break}$ are quite small, see Fig. \ref{fig:specres}), $\rho$ does not significantly depend on $\nu_{break}$. Therefore, in these regions, the effects of the uncertainties in $\nu_{break}$ are also negligible.

\subsection{Step C}

Once $NONTH$ has been produced we simply subtract it from the total image in the oxygen band, thus obtaining the pure thermal image in the $0.5-0.8$ keV band shown in Fig. \ref{fig:thermal}.

\end{document}